
\input harvmac
{\vbox{\centerline{Reply to Comments on ``Asymptotic Estimate of the {\it
n}-Loop}
\vskip2pt\centerline{QCD Contribution to the Total $e^{+}e^{-}$  Annihilation
Cross Section''}}}

\centerline{GEOFFREY B. WEST \footnote{$^\dagger$}{ email: gbw@pion.lanl.gov}}
\centerline{LOS ALAMOS, NM  87545}
\bigskip


The three Comments by Barclay and Maxwell \ref\one{D. T. Barclay and C. J.
Maxwell, Phys. Rev. Letters to be published.}, Duncan and Willey \ref\two{A.
Duncan and R. Willey ibid.} and Samuel and Steinfelds \ref\three{Mark A. Samuel
and Eric Steinfelds, ibid.} all correctly point out that my
estimate \ref\four{G. B. West, ibid {\bf 67}, 1388 (1991).  The notation and
equation numbers in this note all follow those in this reference.} for
$r_n(1)$, the nth order coefficient in the perturbative expansion of the
normalized total $e^{+}e^{-}$ cross-section, does not exhibit the correct
flavor or color dependence of the exact calculation \ref\five{L. R. Surguladze
and M. A. Samuel, Phys. Rev. Lett. {\bf 66}, 560 (1991); S. G. Gorishny, A.
L. Kataev and S. A. Larin, Phys. Lett {\bf B259}, 144 (1991).}.  This
naturally casts serious doubts upon the validity of the result.  I would like
to suggest, however, that, in spite of this, my original estimate remains
valid as an asymptotic formula for sufficiently large {\it n} and that the
leading corrections of $0(1/n)$ have a strong flavor and color dependence.
Below I give some arguments as to why this might be expected to be so.

It turns out that, if the region of validity of the asymptotic formula is
$n\gg n_{0}(n_{f},N_{c})$, then typically $n_0$ is expected to be
relatively large, i.e. $>0(2-3)$.  I propose, however, that there is a valley
in
$(n_{f}, N_{c})$ parameter space whose bottom is the approximate line
$n_{f}\approx 2N_{c}-1$ and where $n_{0}\roughly < 0(1)$.  This passes through,
or near, the point $(n_{f}=5, N_{c}=3)$ so, in this sense, the close agreement
of
my result with the exact calculation for the physical case of interest is,
indeed, fortuitous.  In the general case of arbitrary $n_{f}$ and $N_{c}$ one
would therefore expect to have to go to much larger values of {\it n}
($\roughly > 6$, say) to obtain reasonable agreement.

The original point of my paper was to derive an asymptotic estimate for
$r_{n}$ that gave the correct order of magnitude: specifically to answer the
question ``is $r_3 \sim 5$ or 50?''  This was in response to the confusion
resulting from the incorrect ``exact'' calculation \ref\six{S. Gorishny, A. L.
Kataev and S. A. Larin, Phys. Lett. {\bf B212}, 238 (1988).} which originally
gave $r_{3}\sim 70$.  Even though corrections to my estimate were discussed
in my paper, no serious attempt to evaluate them was made.  Since the
position of the leading saddle point is at $k_{1}\approx b_{1} [(n-1) + b']$,
it was natural to retain the combination $(n+b')$ in the final result.
However as I pointed out, this, of course, does not incorporate all $0(1/n)$
corrections.  There are several other sources of $1/n$ contributions such as
corrections in going from the $d_{n}$ to the $r_{n}$ [see my eq. (12)] and
from approximating $Im D$ by its leading term [see eq. (11)].  The former
leads to contributions to $r_{3}$ like $(8\pi^{2}/3)(b_{2}/b_{1})r_{2}$ and
${(4\pi^{2})}^{2}(b_{3}/3b_{1})$ whereas the latter gives an overall modifying
factor $[1+1/3\{b'+r_{2}/2\pi^{2}b_{1}\}]$.  To these must be added
corrections to the Gaussian approximation of the saddle-point integration.
Although a careful systematic examination of all the corrections has not yet
been carried out (it is presently underway) it is clear that they are, in
general, large and have a strong dependence on $n_{f}$ and $N_{c}$.  For
example, in $\overline{MS}$ with $N_{c}=3$ and $n_{f}=5$, the factor
$r_{2}/2\pi^{2}b_{1}\approx 1.5$, whereas with $N_{c}=5$ and $n_{f}=1$, it is
almost 3.  Furthermore the contribution $(8\pi^{2}/3)(b_{2}/b_{1})r_{2}$ which
is 1.18 for $N_{c}=3, n_{f}=5$, is 14 when $N_{c}=5, N_{f}=1$!  Similarly, the
term $(4\pi^{2})^{2}b_{3}/3b_{1}$ is only 0.5 for $N_{c}=3, n_{f}=5$ but is 7
at $N_{c}=5, n_{f}=1$.  A preliminary evaluation of this set of corrections
indicates that, in general, they can be uncontrollably large in some cases.
However, when $N_{c}=3$ and $n_{f}=5$ they are relatively small so that use of
the asymptotic estimate for $r_{3}$ can be justified.  On the other hand, the
remarkable closeness to the exact result is clearly accidental.

Notice, incidentally, that scheme dependence enters via these non-leading
contributions.  The fact that the leading term is scheme invariant is not an
argument against its validity.  On the contrary, one can argue on very general
grounds that the leading large {\it n} behavior of $r_{n}$ should, in fact, be
scheme invariant.  The point is that this behavior determines the nature of the
divergence of the perturbation series which is itself a reflection of the
singularity structure in $g^{2}$ near $g^{2}=0$.  However, the analytic
structure in $g^{2}$ can be determined via the renormalization group (RG) since
this requires that $q^{2}$ and $g^{2}$ always occur in the combination
$q^{2}e^{K(g)}[{\rm with} K(g)\equiv\int{dg\over\beta(g)}]
$ and the analytic properties in $q^{2}$ are known [see my eq. (9)].  Using the
perturbative expansion for $\beta(g)$ around $g^{2}=0$ gives $K(g)\approx
1/b_{1}g^{2} + b'\ln g^{2} + 0(g^{2})$ when $g^{2}\approx 0$.  The neglected
terms are analytic at $g^{2} = 0$.  The non-analytic structure at $g^{2} = 0$
is
therefore completely determined by $b_{1}$
and $b_{2}$ both of which are scheme invariant.  This therefore shows (i) that
the leading large {\it n} behavior of the $r_{n}$ can, in principle, be
determined from the RG and $q^{2}$ analyticity and (ii) that the result will
depend only on $b_{1}$ and $b_{2}$ and therefore be scheme-invariant.

As a corollary, this also demonstrates the importance of $b_{2}$ since its
presence dramatically changes the analytic structure in $g^{2}$.  From the fact
that there are discontinuities only along the positive real axis in $q^{2}$ one
can deduce [see my eqn. (9)] that the $g^{2}$ (or $k\equiv 1/g^{2}$)
singularities occur when $k/b_{1} + b' \ln (k + {b^{2}\over b_{1}}) + \cdots =
\ln z \pm 2\pi Ni$ with {\it N} an integer and z running from zero to infinity.
When $b_2=0$ this implies that the $k$-plane [where the integration is to be
performed] separates into an infinite number of disconnected sectors parallel
to
the real axis each separated from the next by $2\pi i b_{1}$.  The appropriate
region of integration therefore reduces to $-\infty$ $<$Rek$<$ + $\infty$ and
$0\leq {\rm Imk} \leq 2\pi b_{1}$, the boundary being the integration contour.
This invalidates a derivation of the null result for $d_{n}$ claimed for this
case $(b_{2}=0)$ in \one.  However, it does emphasize a point that was
suppressed in my paper, namely that great care must be taken in defining the
contour and domain of integration as determined by the RG before interchanging
the {\it z} and {\it k} integrals.  Indeed, when $b'=0$, this results in no n!
behavior from the saddle-point integration.  However, if $b'$ is now
included, additional singularities are present, for, when
$z\approx 0$, one can now have $k \approx -b'$.
The structure in the {\it k}-plane is now quite different and leads, via a
saddle-point integration to the {\it n}! growth of $r_{n}$ and the result
quoted in my paper.  The presence of $b_{2}$ is crucial; even though it plays
a minor role in the final expression, the result is simply not deriveable
without it.  Thus, conclusions based on its omission, such as in ref. 1 and in
the work of Brown and Yaffe \ref\seven{L. S. Brown and L. Yaffe, Phys. Rev.
{\bf
45} R398 (1992).  These authors set $b_{2}=0$ and argue that since they cannot
derive my result by their method, then analyticity and the RG are not
sufficient
to determine the large {\it n} behavior of $r_{n}$.  However, they do not
actually use the full analyticity properties of D in the full complex $q^{2}$
plane, only the fact that its absorptive part along the positive real axis is
determined by R.} are not directly relevant.

The problem of interchanging integrations, which requires care in defining the
domain and contour of integration, is nicely illustrated by the example of eq.
(2) in ref. 1.  To avoid the ``zero times infinity'' problem that can occur
from a cavalier interchange, the $g^{2}$-contour in this case should first be
wrapped around the cut associated with $(-g^{2})^{1-s}$ to obtain \ref\eight{G.
B. West, in ''Radiative Corrections'' edited by N. Dombey and F. Bondjema
(Plenum, N.Y., 1990) p. 487.} \eqn\onen{d(s) = \sin \pi s\int_{L}{dg^{2}\over
\pi}(g^{2})^{-1-s}D(g^{2})} with L lying above the cut on the positive real
axis.  This is the appropriate coefficient generating function for this case.
Now, when the representation eq. (1) or ref. 1 is inserted, no problem arises
upon interchange of integrations; thus  \eqn\twon{\eqalign{d(s)&={\sin \pi s
\over \pi}\int^{\infty}_{0}dz f(z)\int^{\infty}_{0} dg^{2}
(g^{2})^{-1-s}e^{-z/g^{2}}\cr  &= \Gamma(s)\{\sin \pi s \int^{\infty}_{0} {dz
\over \pi}f(z)z^{-s}\}.}}
The quantity in curly brackets is just the sth coefficient in the
perturbative expansion of $f(z)$ as it must be.

Many of these points will be expanded upon in greater detail in forthcoming
papers.
\listrefs
\end